\documentclass{aa}
\usepackage{txfonts}
\usepackage{natbib}
\usepackage{graphicx}

\def\src{IGR~J17419$-$2802}

\def\xmm{{\em XMM--Newton}}
\def\sw{{\em Swift}}
\def\inte{{\rm INTEGRAL}}

\def\chandra{{\em Chandra}}

\def \int{{\rm INTEGRAL}}

\def\approxgt{\mathrel{\hbox{\rlap{\lower.55ex \hbox {$\sim$}}
    \kern-.3em \raise.4ex \hbox{$>$}}}}
\def\approxlt{\mathrel{\hbox{\rlap{\lower.55ex \hbox {$\sim$}}
    \kern-.3em \raise.4ex \hbox{$<$}}}}

\def\ltsima{$\; \buildrel < \over \sim \;$}
\def\lsim{\lower.5ex\hbox{\ltsima}}
\def\gtsima{$\; \buildrel > \over \sim \;$}
\def\gsim{\lower.5ex\hbox{\gtsima}}

\def\hcm {\hbox {\ifmmode $ atom cm$^{-2}\else atom cm$^{-2}$\fi}}
\def\arcmin {\hbox{$^\prime$}}

\def \apjl {ApJL}
\def \apjs {ApJS}
\def \aap {A\&A}

\newcommand{\be}{\begin{equation}}

\newcommand{\ee}{\end{equation}}

\newcommand{\swift}{{\emph{Swift}}}

\begin{document}
\title{Unveiling the short and faint X-ray transient nature of IGR~J17419$-$2802}

\author{V.~Sguera\inst{1},  L.~Sidoli\inst{2} 
}
\institute{
$^1$ INAF$-$OAS, Osservatorio di Astrofisica e Scienza dello Spazio, Area della Ricerca del CNR, via Gobetti 101, I-1-40129 Bologna, Italy \\
$^2$ INAF$-$IASF, Istituto di Astrofisica Spaziale e Fisica Cosmica, via A.\ Corti 12, 20133 Milano, Italy \\
}

\offprints{V. Sguera, vito.sguera@inaf.it}

\date{Received 23 May 2024 / Accepted 27 August 2024}

\authorrunning{V. Sguera et al.}

\titlerunning{Short X-ray transient IGR~J17419$-$2802}

\abstract{We report new X-ray results from the INTErnational Gamma-Ray Astrophysics Laboratory (\inte), \sw, \chandra, and \xmm\ observations of the hitherto poorly studied unidentified X-ray transient \src. We studied in detail the temporal, spectral, and energetic properties of three hard X-ray outbursts detected above 20 keV by \inte. They are all characterized by an average X-ray luminosity of 3$\times$10$^{35}$~erg~s$^{-1}$ and a constrained duration of a few days. This marks a peculiarly short and faint X-ray transient nature for \src. From archival unpublished soft X-ray observations, we found that the source spends most of the time undetected at very low X-ray fluxes (down to $<4.7\times10^{-14}$ erg cm $^{-2}$ s$^{-1}$) for a dynamic range $>$2,000 when in outburst. We provided an accurate arcsecond-sized source error circle. Inside it, we pinpointed the best candidate near-infrared counterpart whose photometric properties are compatible with a late-type spectral nature. 
Based on our new findings, we suggest that \src\ is a new member of the very faint X-ray transients (VFXTs) class. Detailed investigations of VFXT outbursts above 20 keV are particularly rare. In this respect, our reported \inte\ outbursts are among the best studied to date; in particular, their constrained duration of a few days is among the shortest ever measured for a VFXT.
}

\maketitle

\section{Introduction}
IGR~J17419$-$2802 is an unidentified X-ray transient discovered by \int\ above 20 keV on 29 September 2005 during observations of the Galactic centre region (Grebenev et al. 2005). A \swift/X-Ray Telescope (XRT) follow-up, performed a few days later, detected the soft X-ray counterpart (Kennea et al. 2005). Renewed hard X-ray activity was detected by \int\ on 19 February 2006 (Grebenev et al. 2006); a radio follow-up with the Very Large Array did not detect any activity at 4.86 GHz (Rupen et al. 2006). All the above information has been reported in brief communications (astronomer's telegrams), which contain no details on the source characteristics, that is, the temporal duration and energetics of the outbursts, and spectral behaviour. IGR~J17419$-$2802 is a poorly studied unidentified X-ray transient whose characteristics are largely unknown. 

Here, we report new X-ray results obtained from unpublished \swift/XRT, \chandra, \xmm, and \int\ archival X-ray observations of the source. In addition, we performed a detailed spectral and temporal investigation of the \int\ data pertaining to the source discovery in 2005 and its renewed activity in 2006. We investigated the characteristics and the nature of the pinpointed lower energy counterpart using optical/infrared archival data. Finally, we used all of the newly reported results to explore the possible nature of IGR~J17419$-$2802.

\section{INTEGRAL results}

\subsection{Observations and data reduction}

The temporal and spectral behaviour of \src\ has been investigated in detail above 20 keV with the INTEGRAL Soft Gamma-Ray Imager (ISGRI) detector (Lebrun et al. 2003), which is the lower energy layer of the  Imager on Board the INTEGRAL Satellite (IBIS) coded mask telescope (Ubertini et al. 2003) on board \int~(Winkler et al. 2003). \textrm{INTEGRAL} observations are divided into short pointings, that is, Science Windows (ScWs), whose typical duration is $\sim$ 2,000 seconds. We specifically searched the IBIS/ISGRI public data archive (from revolution 30 to 2,000, i.e. from approximately January 2003 to September 2018) for hard X-ray activity from \src\. In particular, the data set consists of 19,204 ScWs where the source was within the instrument field of view (FoV) with an off-axis angle value less than 12$^{\circ}$. This limit is generally applied because the response of IBIS/ISGRI is not well modelled at large off-axis values, and this may introduce a systematic error in the measurement of the source fluxes. The effective exposure time is equal to 27.8 Ms. IBIS/ISGRI flux maps for each ScW were generated in different energy ranges (i.e. 18--60 keV, 17--30 keV, and 20--40 keV) using the latest release 11.2 of the Offline Scientific Analysis software (OSA). Count rates at the position of the source were extracted from individual ScW flux maps, to build long-term light curves at the ScW level, that is, with a bin time of $\sim$ 2,000 s.

To search for transient hard X-ray activity (E$>$20 keV) from \src~in a systematic way, we used the bursticity method as developed by 
Bird et al. (2010, 2016). This method optimizes the source detection timescale by scanning the IBIS/ISGRI light curves with a variable-sized time window to search for the best source significance value on timescales ranging from 0.5 days to 
months or years. Then, the duration, time interval, and energy band over which the source significance is maximized were recorded. Once a newly discovered outburst from \src~was found using the bursticity method, an in-depth spectral and temporal analysis was performed. To this aim, the data reduction was carried out using the OSA 11.2 software. The IBIS/ISGRI systematics, which are typically of the order of 1$\%$, were added to the extracted spectra.

Using the bursticity method, our investigation of the IBIS/ISGRI data archive (2003--2018) allowed us to find three hard X-ray outbursts best detected in the energy band 18--60 keV. They were detected in September 2005, February 2006, and April 2006, respectively. Table 1 reports a summary of their characteristics. We note that the detection of outbursts n. 1 and n. 2 has already been reported in the literature, albeit through very brief communications that contain no information on their spectral and temporal characteristics (Grebenev et al. 2005, 2006). Conversely, outburst n. 3 is newly discovered and reported for the first time in this work. 

\src\ is not detected as a persistent source in the latest published \int/IBIS catalogue (Bird et al. 2016), despite an extensive coverage of its sky region ($\sim$11 Ms). This information can be used to infer an upper limit on its persistent hard X-ray emission. To estimate the statistical upper limit, the variance value was extracted at the source position from the global mosaic variance map once all of the outbursts were excluded.
The inferred 3$\sigma$ upper limit is equal to $\sim$0.2 mCrab, or 1.6$\times$10$^{-12}$ erg cm$^{-2}$ s$^{-1}$ (20--40 keV). 

In the following, we report a comprehensive \int\ study of the temporal, spectral, and energetic characteristics of the three hard X-ray outbursts reported in Table 1. 

 \subsection{Outburst n. 1}
 
 \begin{table}
 \caption {Characteristics of the three outbursts studied in this work (duration, \int\ revolutions and their start date, effective exposure time on source, significance of the source detection, and average 18--60 keV flux).} 
\label{tab:main_outbursts} 
\begin{tabular}{ccccccc}
\hline
\hline  
  out         & dur  &rev     &   start time       & exp          &  sig                     &  flux         \\
  (n.)     & (days)   &(n.)     & (MJD)              &  (ks)        &  ($\sigma$)             &  (mCrab)       \\   
\hline 
        &        &359+360    &  53636.028       &   22       &     $<$2$\sigma$          &  $\le$ 2.1           \\ 
             
 1  & $\sim$4   &361       &  53640.384    &    19    &    6.5$\sigma$         &  7.2$\pm$1.1               \\
   &         &362       &   53643.385      &    34    &    11.7$\sigma$         &  9.6$\pm$0.8                \\
 
   &        &363       &   53646.373       &    29     &    $<$2$\sigma$             &  $\le$ 1.8             \\

 \hline
   &    &408      &   53780.895       &   31       &     $<$2$\sigma$          &  $\le$ 1.8          \\ 
2  &  $\sim$2.5   &409       &   53783.882       &    40     &    11$\sigma$            &  8.3$\pm$0.8           \\
   &  & 410+411   &  53786.869       &    27     &     $<$2$\sigma$       &   $\le$ 1.9              \\     

\hline 
   &         &424      &   53828.701       &   47       &     $<$2$\sigma$          &  $\le$ 1.4          \\ 
3  & $\sim$5.5   &425       &  53831.688    &    46    &    5$\sigma$         &  3.4$\pm$0.7               \\
   &         &426       &   53834.677      &    43    &    11.5$\sigma$         &  8.2$\pm$0.7                \\
   &         &427      &53837.664&   35   &     $<$2$\sigma$          &  $\le$ 1.6         \\ 
 \hline
\hline  
\end{tabular}
\end{table}

\begin{figure}
\begin{center}
\includegraphics[height=8.5 cm, angle=270]{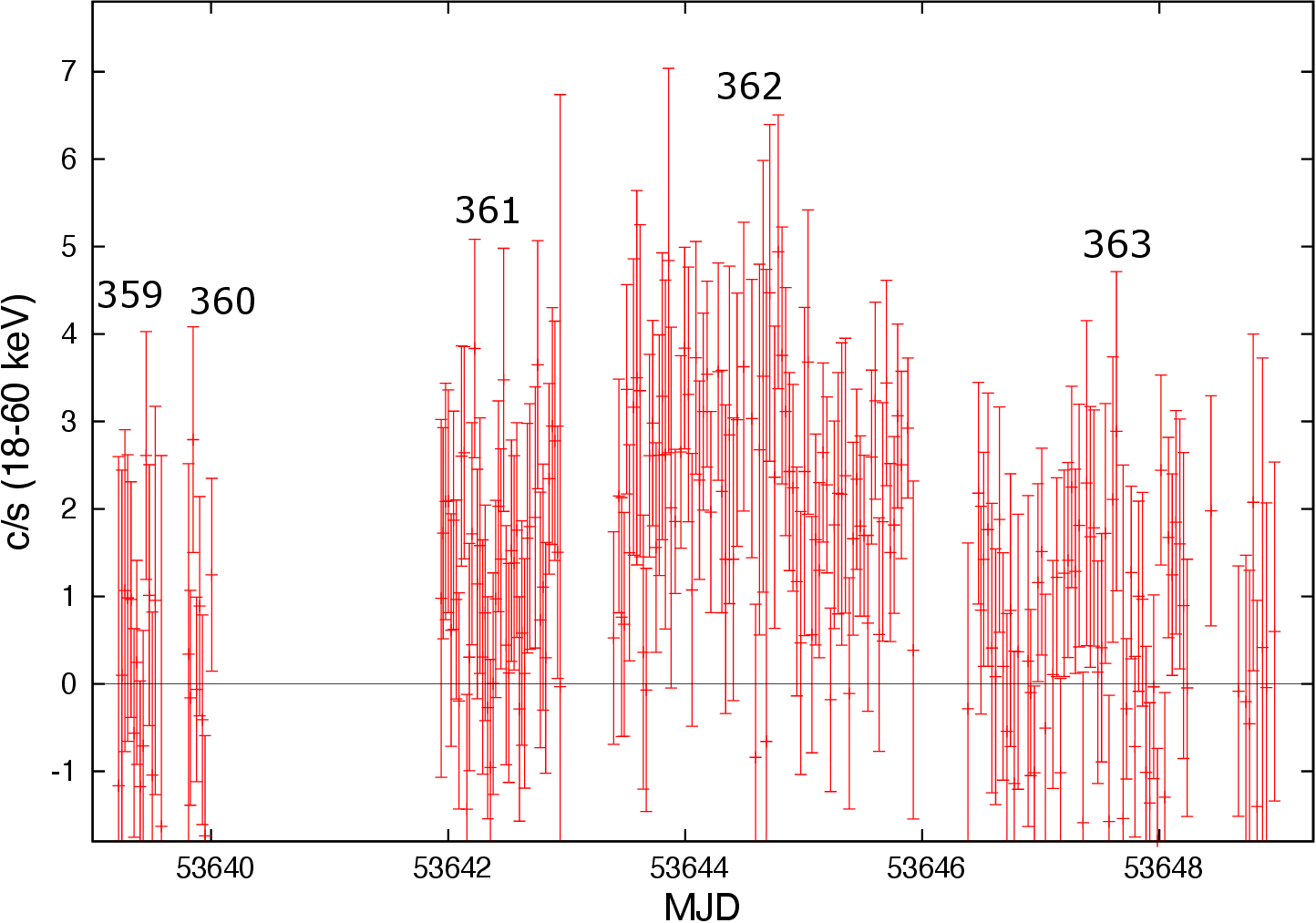}
\caption{IBIS/ISGRI light curve (18--60 keV) from revolution 359 to 363 of outburst n. 1 in Table 1. The bin time is at the ScW level, i.e. $\sim$ 2,000 s.}
\end{center}
\end{figure}

\begin{figure}
\begin{center}
\includegraphics[height=6 cm, angle=0]{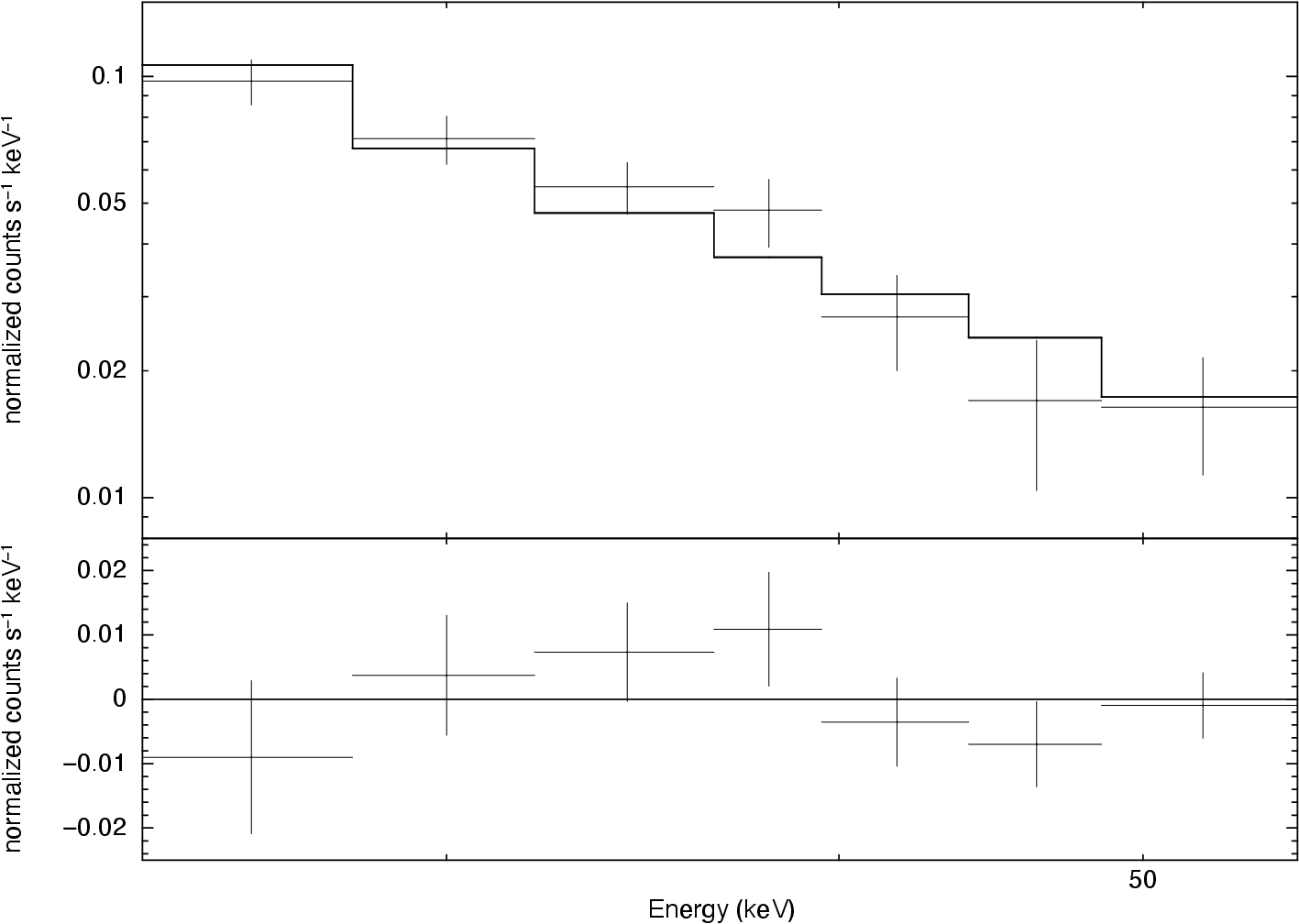}
\caption{IBIS/ISGRI spectrum during outburst n. 1 (extracted from revolution 361+362). It shows the power law data-to-model fit with the corresponding residuals.}
\end{center}
\end{figure}

 \src\ was detected during revolution 361 starting on 28 September 2005 at 22:42 UTC (see Table 1). Since the source was not significantly detected during the previous revolutions, we can safely assume this date as the starting time of the outburst activity. Similarly, we can consider the ending date of revolution 362 (2 October 2005 at 22:45 UTC) as the ending time of the outburst activity, since \src\ was not significantly detected during the subsequent revolution. Hence, the duration of the outburst is 
$\sim$4 days. The outburst was best detected in the energy band 18--60 keV, no detection was obtained in the higher 
energy band 60--100 keV, and the inferred 2$\sigma$ upper limit is 4.3 mCrab.

 The IBIS/ISGRI light curve at the ScW level (Fig. 1) clearly shows, in its middle, the outburst activity detected during revolutions 361 and 362. The measured peak flux (at MJD 53644.8) is 25$\pm$7 mCrab, or $\sim$3.2$\times$10$^{-10}$ erg cm$^{-2}$ s$^{-1}$ (18--60 keV), which is higher than the average flux by a factor of $\sim$3. 
The source was never significantly detected (i.e. $\ge$ 5$\sigma$) at the ScW level at any point in the entire outburst.

We extracted the average IBIS/ISGRI spectrum of the outburst during revolutions 361 and 362. 
The best fit is achieved with a power law model ($\chi^{2}_{\nu}$=0.98, 6 d.o.f.) with photon index $\Gamma$=2.8$\pm$0.5. 
The average 18--60 keV (20--40 keV) flux is 1.1$\times$10$^{-10}$ erg cm$^{-2}$ s$^{-1}$ (6.6$\times$10$^{-11}$ erg cm$^{-2}$ s$^{-1}$). 
Alternatively, a thermal bremsstrahlung model provides a reasonable fit ($\chi^{2}_{\nu}$=0.78, 6 d.o.f.) with  kT=24$^{+11}_{-6}$ keV. 
Fig. 2 shows the power law data-to-model fit with the corresponding residuals.

Images from the Joint European Monitor for X rays (JEM--X), the X-ray monitor on board  \int\  (Lund et al. 2003), were created in two energy bands (3--10 keV and 10--20 keV) for the reported outburst. The source was also in the JEM--X1 FoV for an effective exposure time of $\sim$47 ks.
\src\ was weakly detected at 6$\sigma$ (5.5$\sigma$) in the energy band 3--10 keV (10--20 keV) with an average flux of 6.20$\pm$0.95 mCrab, or $\sim$9.2$\times$10$^{-11}$ erg cm$^{-2}$ s$^{-1}$ (8.1$\pm$1.5 mCrab, or $\sim$6.5$\times$10$^{-11}$ erg cm$^{-2}$ s$^{-1}$). 
The statistics are insufficient for performing any meaningful spectral or temporal analysis.

 \subsection{Outburst n. 2} 
From Table 1 we note that \src\ has been detected only during revolution 409, which started on 17 February 2006 at 21:11 UTC and ended on
20 February 2006 at 10:29 UTC. Hence, the inferred outburst duration is $\sim$2.5 days. 
Fig. 3 shows the IBIS/ISGRI light curve at the ScW level. We note that the source was significantly detected at the ScW level (i.e. $\ge$ 5$\sigma$) 
only during one pointing (starting on 20 February 2006 at 02:31 UTC, or MJD 53786.1), during which it reached a 18--60 keV peak flux of 26.1$\pm$4.5 mCrab, or $\sim$3.2$\times$10$^{-10}$ erg cm$^{-2}$ s$^{-1}$ (similarly to outburst n. 1).

 \begin{figure}
\begin{center}
\includegraphics[height=6 cm, angle=0]{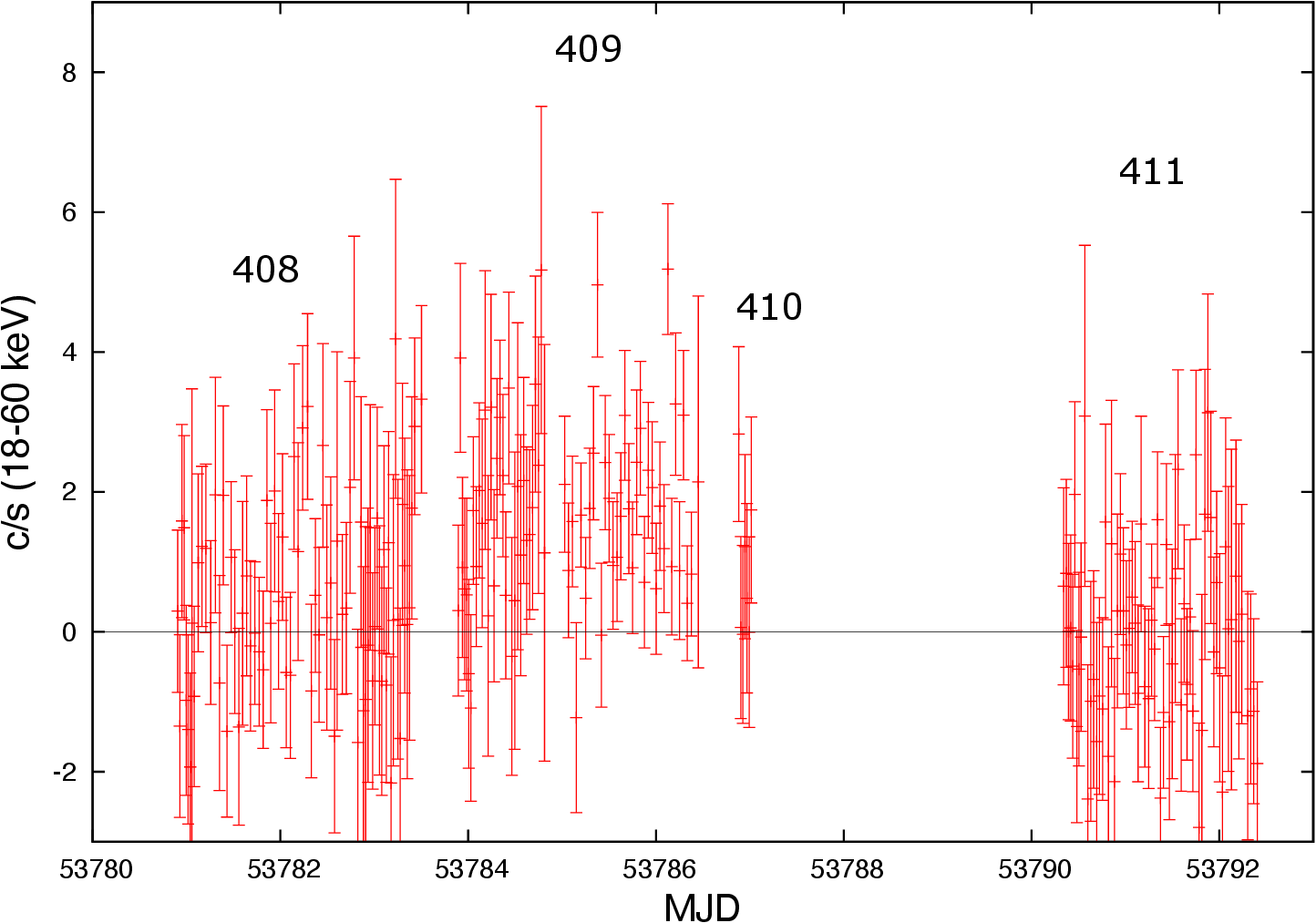}
\caption{IBIS/ISGRI light curve (18--60 keV) from revolution 408 to 411 of outburst n. 2 in Table 1. The bin time is at the ScW level, i.e. $\sim$ 2,000 s.}
\end{center}
\end{figure}

\subsection{Outburst n. 3} 
This is a newly discovered outburst, never reported to date. From Table 1, we note that the hard X-ray activity started during revolution 425 on 6 April 2006 at 16:30 UTC and ended during revolution 426 on 12 April at 05:35 UTC. Its duration is $\sim$5.5 days. Fig. 4 shows the IBIS/ISGRI light curve.

 \begin{figure}
\begin{center}
\includegraphics[height=6 cm, angle=0]{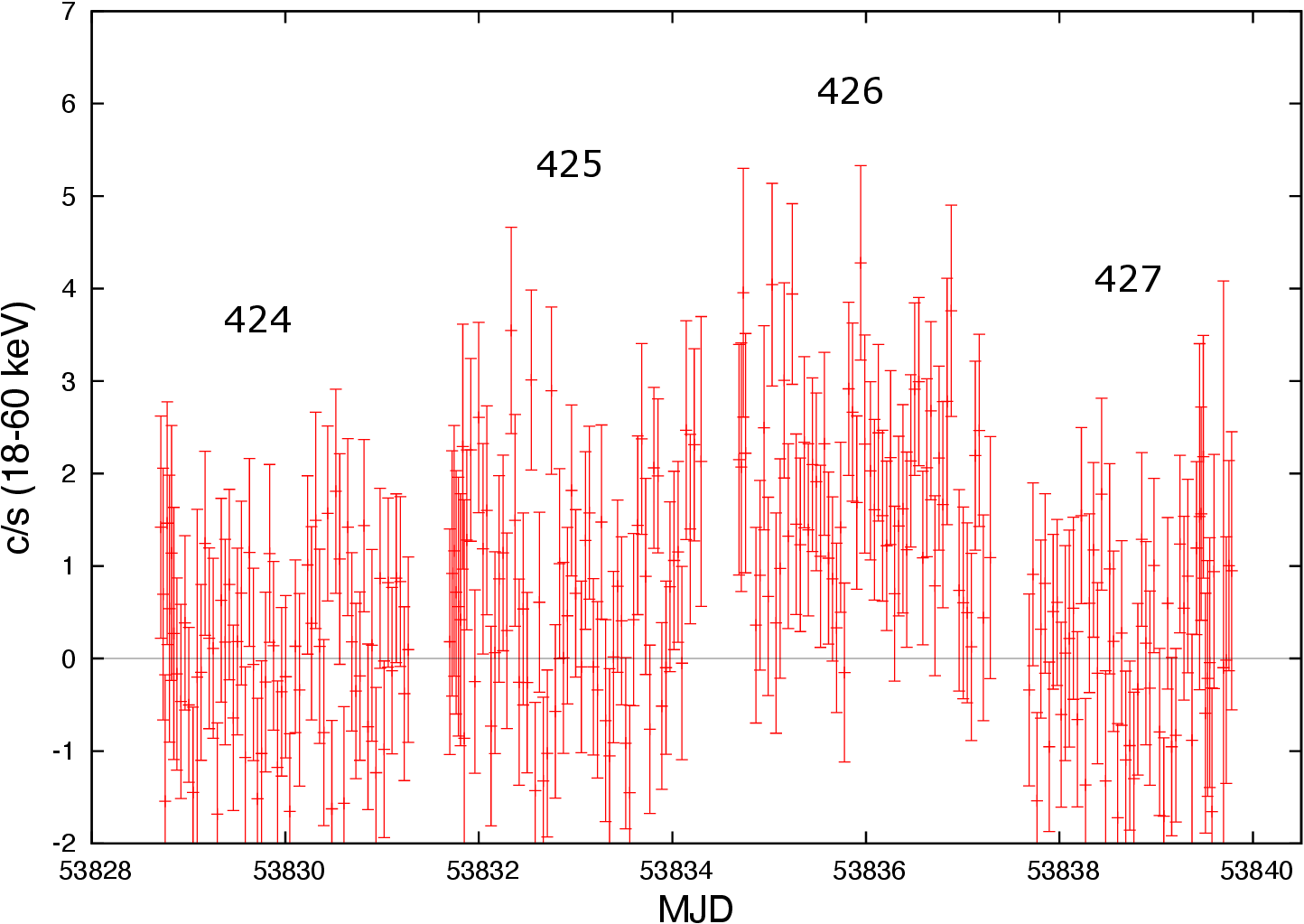}
\caption{IBIS/ISGRI light curve (18--60 keV) from rev 424 to 427 of outburst n. 3 in Table 1. The bin time is at the ScW level, i.e. $\sim$ 2,000 s.}
\end{center}
\end{figure} 


 \section{Soft X-ray observations and results}
\label{sect:data}

In the following, we report the results obtained in the soft X--ray band
using archival data from different space missions (see details in Table A.1).
 When a meaningful spectroscopy was possible, we adopted the interstellar abundances from Wilms et al. (2000) and the photoelectric absorption cross-sections from Verner et al. (1996). All spectral uncertainties are given at a 90\% confidence level for one parameter of interest.

\subsection{\chandra}
\label{sect:chandra}

\chandra\ observations covered the source position seven times using the Advanced CCD Imaging Spectrometer 
(ACIS-I)  instrument (0.5-7 keV). 
In Table A.1, we report the summary of these observations, together with their nominal exposure times. 
The data were reprocessed and analysed with standard procedures 
using the \chandra\ Interactive Analysis of Observation (CIAO 4.14) and CALDB (4.9.8).
The CIAO tool {\sc fluximage} was used to build sky images and exposure maps (0.5-7.0 keV) and to estimate the radii of
the circular regions enclosing 90\% of the point spread function (PSF) at 1.0 keV.
These radii are between 0.9 arcsec (on-axis position) to 15.5 arcsec for the most off-axis position. 
Annular regions centred on the \swift-XRT source position (inner and outer radii of one and five times the
source extraction radius) were used to estimate the contribution of the background.
The effective exposures used for the flux computation take into account the off-axis positions of the sources.
Since no X-ray sources were detected at the source position, $3\sigma$ upper limits 
to the count rate were estimated in the 0.5-7 keV energy band (Kraft et al. 1991).
These values were then converted to unabsorbed fluxes (0.3-10 keV, to compare with \swift\ results) 
using calibration files appropriate for the observation dates and off-axis positions.
A power law model was assumed, with a photon index $\Gamma$=2 and 
Galactic absorption N$_{\rm H}=1\times10^{22}$cm$^{-2}$.
The results are reported in Fig. 5 and in Table A.1.

\begin{figure}
\begin{center}
\includegraphics[width=9cm,angle=0]{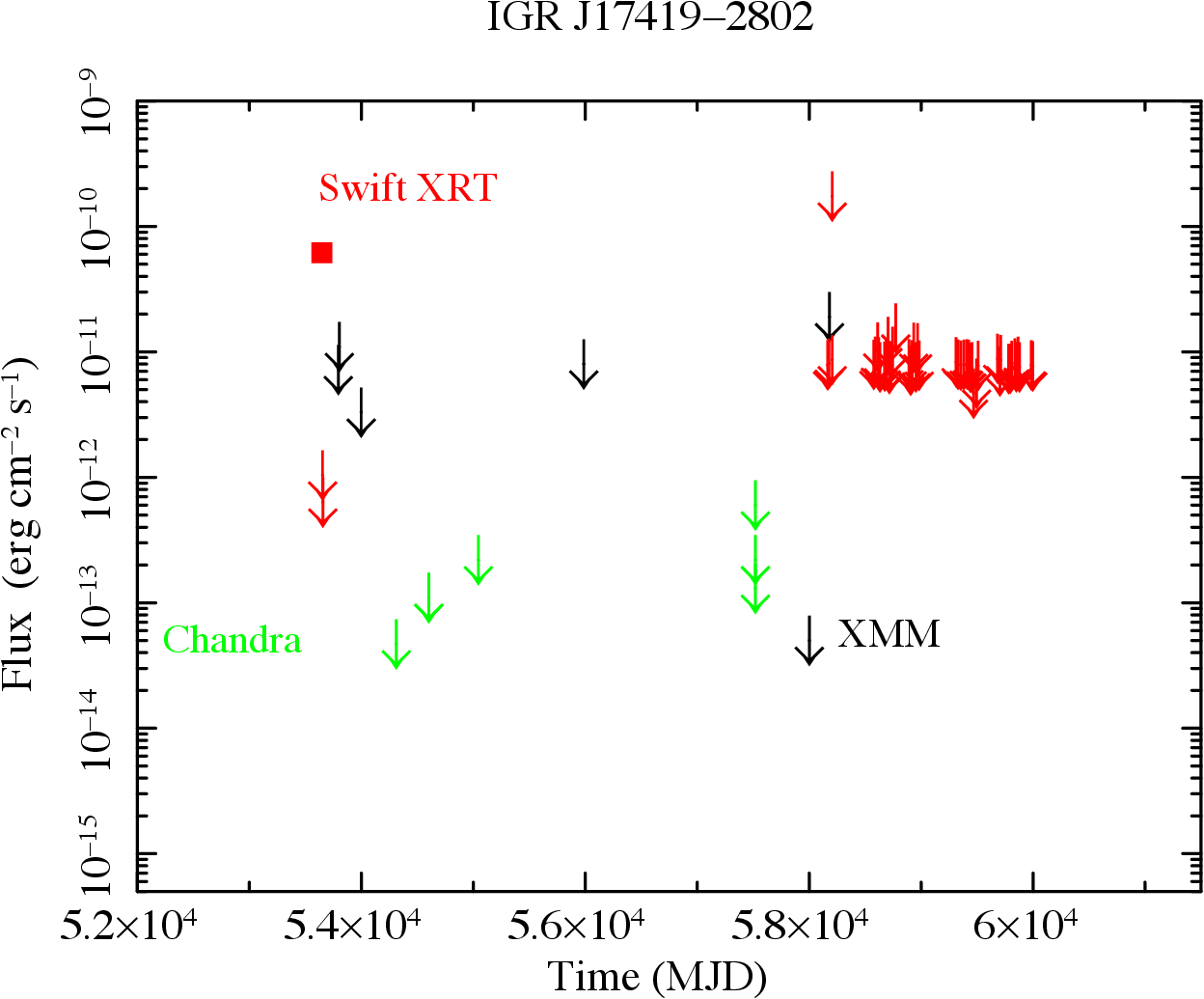}
\caption{Long-term light curve (0.3–10 keV) of IGR J17419$-$2802 obtained with Swift-XRT, \xmm, and \chandra\ data. 
The unabsorbed fluxes (both detections and $3\sigma$ upper limits) are taken from Table A.1.}
\label{fig:lc}
\end{center}
\end{figure}

\subsection{\xmm}
\label{sect:xmm}

The \xmm\ satellite (Jansen et al. 2001) covered the source sky position with  the European Photon Imaging Camera
(EPIC; Struder et al. 2001, Turner et al. 2001) 
during a pointed observation only once, on 5 September 2017 (ObsID 0801680101; T$_{exp}$=28 ks), 
at a large off-axis angle of $\sim$10 arcmin.
We reduced the data using the \xmm\ Science Analysis Software ({\sc sas}, version 21.0.0), adopting standard procedures.
The source was undetected below a flux corrected for the absorption of 
UF $<5.0\times10^{-14}$~erg cm$^{-2}$ s$^{-1}$ (0.3-10 keV, assuming a power law model with a photon index $\Gamma$=2; see below).
The source location was also serendipitously covered during short slew manoeuvres that 
reorientered the spacecraft between targets. 
Here, 3$\sigma$ upper limits on its X-ray flux from both the 2017 observation and the slew manoeuvres
were obtained from the ESA HIgh energy LIghtcurve GeneraTor
(HILIGT\footnote{http://xmmuls.esac.esa.int/upperlimitserver}; Saxton et al. 2022, Konig et al. 2022).
HILIGT returns upper limits on absorbed fluxes (0.2-12 keV) assuming a power law model with
a photon index $\Gamma$=2 and an absorbing column density N$_{\rm H} = 1\times10^{21}$cm$^{-2}$.
We converted these values to the 0.3-10 keV energy range
assuming N$_{\rm H}$=1$\times10^{22}$cm$^{-2}$ and a power law model with $\Gamma$=2. 
These values are reported in Table A.1 and shown in Fig. 5.


\subsection{The \textit{Neil Gehrels} \swift\ Observatory}
\label{sect:swift}

\begin{figure*}
 \includegraphics[width=8cm, angle=0]{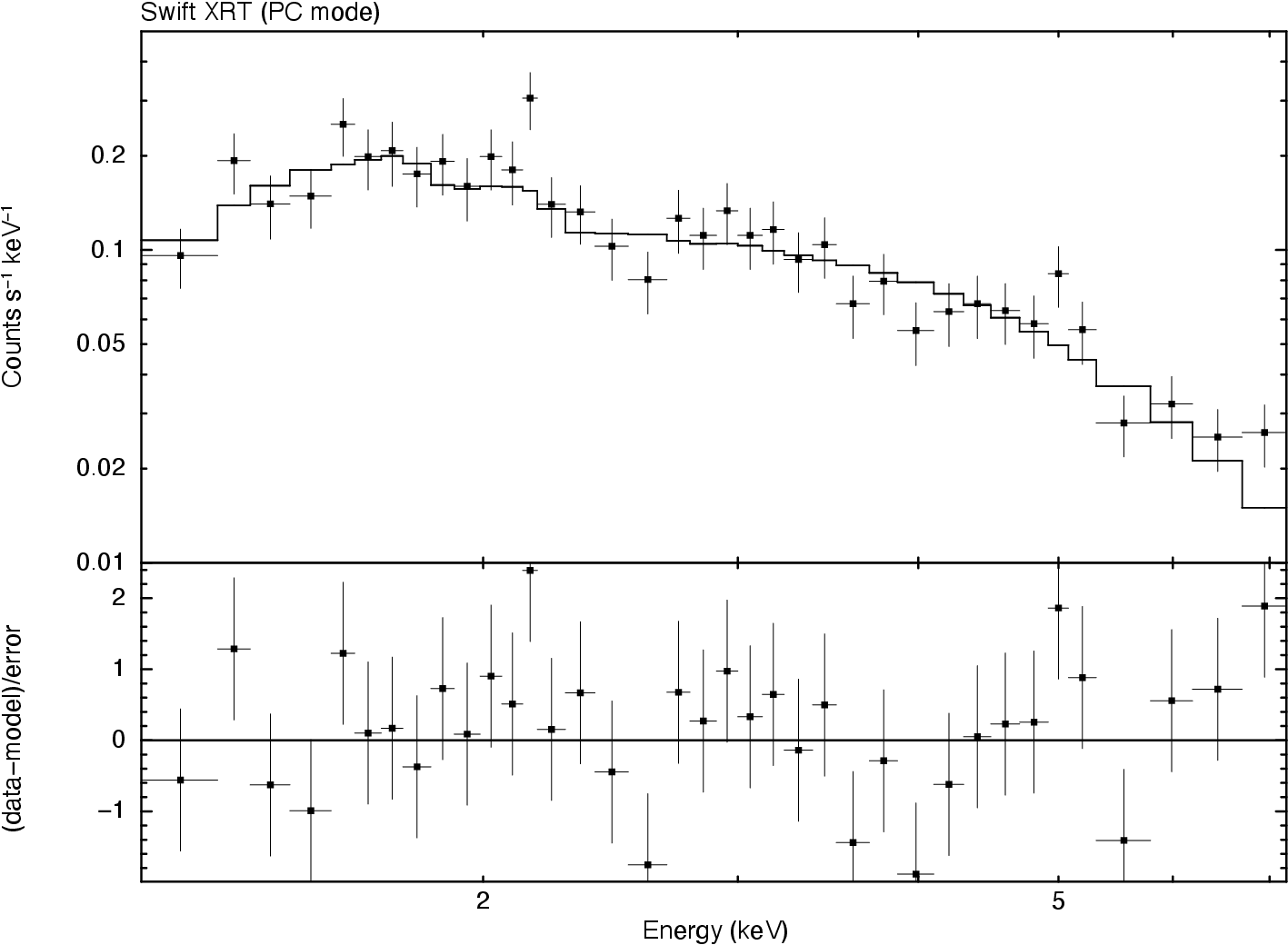} 
 \includegraphics[width=8cm, angle=0]{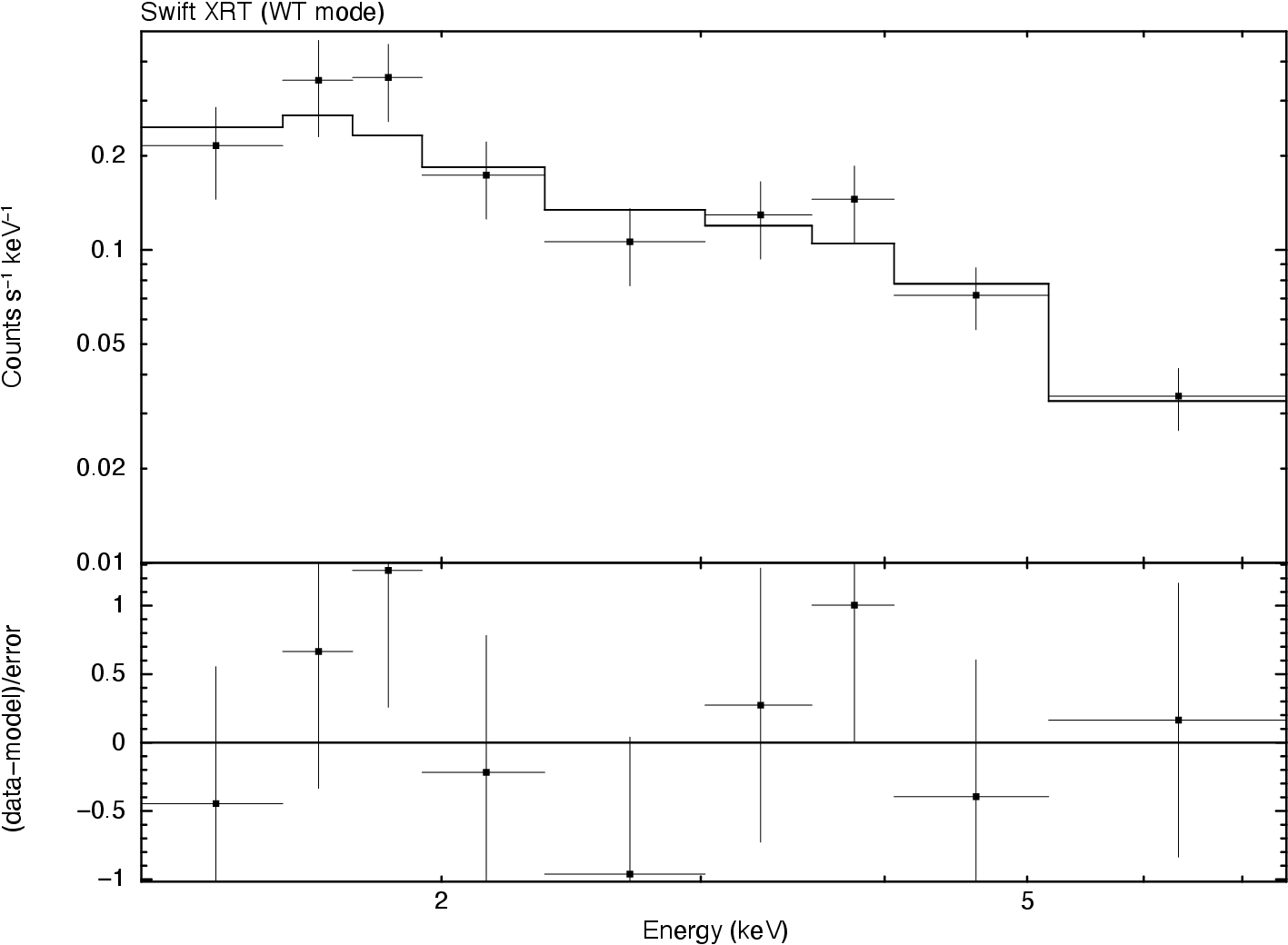}
\caption{\swift-XRT spectra obtained from ObsID 30330001, in PC (\textit{left panel}) 
and WT (\textit{right panel}) data modes.
An absorbed power law model was adopted. Count spectra together with
the residuals in units of standard deviations are shown. 
}
  \label{fig:xrtspec}
\end{figure*}

\begin{table*}
	\centering
	\caption{\src\ \swift-XRT spectra (ObsID~30330001) fitted with an absorbed power law model.
	Fluxes are in the 1-10 keV energy range.}
	\label{tab:xrtspec}
	\begin{tabular}{lccc} 
		\hline
	Param.	             &  PC          &    WT   & PC \& WT \\	
		\hline
  N$_{\rm H}$ (10$^{22}$~cm$^{-2}$) & 2.1 $^{+0.6} _{-0.5}$ & 1.4 $^{+1.9} _{-1.4}$  & 1.8 $^{+0.7} _{-0.6}$ \\     	
  $\Gamma$              & 1.7 $^{+0.3} _{-0.3}$ & 1.5 $^{+0.7} _{-0.7}$  & 1.6 $^{+0.4} _{-0.3}$ \\
  Obs. Flux (erg cm$^{-2}$ s$^{-1}$)         & 4.4$\times10^{-11}$  & 5.3$\times10^{-11}$ &  4.8$\times10^{-11}$ \\  
Unabs. Flux (erg cm$^{-2}$ s$^{-1}$

)          & 5.9$^{+0.6} _{-0.5}\times10^{-11}$ & 6.3$^{+1.9} _{-1.2}\times10^{-11}$  & 6.2$^{+1.1} _{-1.0}\times10^{-11}$ \\ 
  $\chi^{2}$ (dof)           & 35.22 (34)       &  4.63 (6)  & 39.34 (41) \\
 \hline
	\end{tabular}
\end{table*}

The \textit{Neil Gehrels} \swift\ Observatory (Gehrels et al. 2004) 
pointed the X-Ray Telescope (XRT; Burrows et al. 2005) 
at the source position three times (Table A.1) during target of opportunity observations  triggered by the source discovery with \int\ on 29 September 2005. 
We downloaded the data from the High Energy Astrophysics Science Archive Research Center (HEASARC) and reduced them with standard procedures
using the {\sc xrtpipeline} tool available in {\sc heasoft} v.6.32.
The source was detected only during ObsID 30330001 (3 October 2005), with the data mode 
switching from photon counting (PC, T$_{exp}$=1550~s) to windowed timing (WT) for 250~s. 
During the other two pointings (11 and 13 October ), the source was not detected and the relative upper limits were obtained by means of 
the UK \swift\ Science Data Centre (UKSSDC\footnote{https://www.swift.ac.uk/}; Evans et al. 2020; see below).
These upper limits provide strong evidence of the short duration of the outburst X-ray activity. 

In addition to the \swift-XRT targeted pointings reported above, there are several short observations
serendipitously covering the source sky position with \swift (Table A.1) We made use of the UKSSDC (Evans et al. 2020) to download the 3$\sigma$ upper limits on the source net count rates (XRT/PC; 0.3-10 keV), per-observation binning, calculated adopting the Bayesian approach by Kraft et al. (1991).
To obtain the long-term light curve shown in Fig. 5,
we converted the upper limits on the XRT/PC rates to unabsorbed fluxes (0.3-10 keV),
adopting an average factor of 1.$\times$10$^{-10}$ erg cm$^{-2}$ count$^{-1}$
(appropriate for a power law model with $\Gamma$=2 and N$_{\rm H}=1\times10^{22}$cm$^{-2}$).
The results are listed in Table A.1.

Pertaining to the \swift-XRT detection on 3 October 2005, 
meaningful X-ray spectra were extracted from both data modes covering the 1-7 keV energy range, 
resulting in net count rates of (4.96$\pm{0.18}$)$\times10^{-1}$~count~s$^{-1}$ (XRT/PC) and
(6.45$\pm{0.59}$)$\times10^{-1}$~count~s$^{-1}$ (XRT/WT). 
The two X-ray spectra appear featureless and are well fitted by an absorbed power law model (Fig. 6).
The spectral parameters are reported in Table 2, where we also report on the joint fit of WT and PC mode spectra.
In particular, we note that when detected, the source displayed a relatively low unabsorbed flux ($\sim$6$\times10^{-11}$~erg~cm$^{-2}$~s$^{-1}$), which 
clearly indicates a rapid fading observed very few days after the discovery by \int. 
In addition, we performed a joint, although not strictly simultaneous, spectral fit (0.3--60 keV) to the \swift/XRT (both WT and PC mode spectra) and IBIS/ISGRI data of outburst n. 1. A multiplicative factor for each instrument was included in the fit to take into account the uncertainty in the cross-calibration of the instruments. Firstly, we used an absorbed power law model, which resulted in a bad description of the broadband data ($\chi^{2}_{\nu}$=1.2, 48 d.o.f.). The fit was significantly improved by considering an absorbed cutoff power law ($\chi^{2}_{\nu}$=0.99, 47 d.o.f.); 
the fit parameters are $\Gamma$=1.5$\pm$0.3, N$_{\rm H}$=(1.7$\pm$0.7)$\times$10$^{22}$~cm$^{-2}$, and a high energy cutoff of 26$^{+25}_{-9}$ keV. The observed broadband flux (0.3--60 keV) is equal to 3$\times10^{-10}$~erg~cm$^{-2}$~s$^{-1}$. Figure 7 shows the unfolded broadband X-ray spectrum. We note that this is the typical phenomenological description model of accreting X-ray pulsars, usually characterized by an absorbed flat power law ($\Gamma$ between 0 and 1) modified by a high energy cutoff in the 10-30 keV range (e.g. Kretschmar et al. 2019). We note that although the intercalibration constant between Swift and \int/IBIS is expected to be around 1, we found a smaller value, equal to 0.3$^{+0.2}_{-0.1}$. In principle, this should imply some variability in the source flux between the observations, since they are not strictly simultaneous. Indeed, the \int\ observation covered the initial brightest part of the outburst activity, while the \swift/XRT observation covered the final and weaker part of it when the source was significantly fading.

\begin{figure}
\begin{center}
\includegraphics[height=8.5 cm, angle=270]{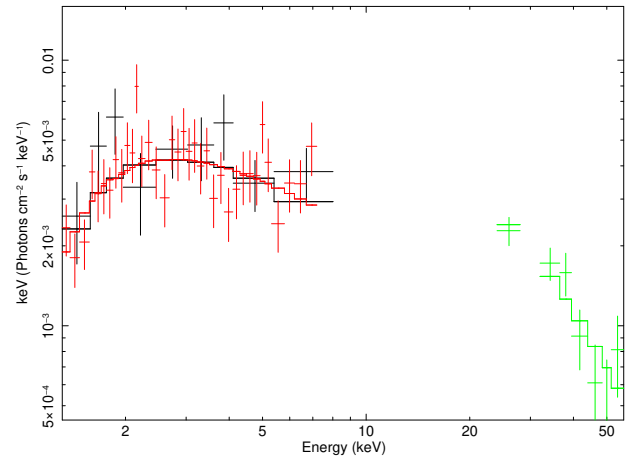}
\caption{Unfolded combined \swift-XRT (WT and PC mode) and IBIS/ISGRI spectrum of \src\ (0.3--60 keV) during outburst n. 1 in Table 1.}
\end{center}
\end{figure}

From the \swift-XRT detection on 3 October 2005, we obtained the best determined XRT position 
at R.A. (J2000) = $17^{\rm h}41^{\rm m}55\fs76$ Dec (J2000) = $-28^\circ01\arcmin55\farcs4$ with a 90$\%$ confidence (99$\%$) error radius of 1$''$.8 (2$''$.55) using the XRT--UVOT alignment and matching UVOT field sources to the USNO-B1 catalogue\footnote{http:/www.swift.ac.uk/user$\textunderscore$objects} (see Evans et al. 2009). We note that our position 
is significantly more accurate and refined than those reported in Kennea et al. (2005) and Kong (2006), which were obtained using the standard procedure and have 90$\%$ confidence radii of 8$''$ and 3$''$.6, respectively.

{As a final note, the long-term soft X-ray source light curve and the unabsorbed fluxes reported in Table A.1 were obtained using a power law model ($\Gamma$=2 and N$_{\rm H}=1\times10^{22}$cm$^{-2}$) with spectral parameters well within the 3$\sigma$ uncertainty of the \swift\ results (PC and WT modes; Table 2). 
However, if we adopt the \swift\ best fit reported in Table 2
($\Gamma$=1.6 and N$_{\rm H}=1.8\times10^{22}$cm$^{-2}$), all soft X-ray unabsorbed fluxes increase by 20\%, at most, depending on the space mission. 
This is irrelevant for the aim of this paper.

\section{Search for infrared and optical counterparts}

\begin{table*}
\caption {NIR sources (as taken from the UKIDSS Galactic Plane Survey) located inside the 90\% confidence error circle (n. 1) and 99\% confidence error circle (from n. 2 to n. 4) of IGR J17419$-$2802.}
\begin{tabular}{ccccccccc}
\hline
\hline  
n.  & name  &  J     &  H     &    K     & offset   & Q & Spectral Type & d \\
   &    &  (mag)   &  (mag)   &    (mag)   &        &  &       & (kpc) \\
\hline  
1   & J174155.70-280155.7  &   17.655$\pm$0.109   &   $>$19.0   &  $>$19.0     &  0$''$.72  &  &  &  \\ 
\hline 
2  & J174155.90-280153.9  & 17.908$\pm$0.138 & 16.837$\pm$0.128 &  16.674$\pm$0.187 & 2$''$.46 & 0.79 &  late type &  \\
3  & 	J174155.71-280153.0 &  15.846$\pm$0.021  & 14.563$\pm$0.016  &  14.143$\pm$0.018  & 2$''$.46 & 0.57  & late type &  \\
4  & J174155.79-280157.9  & 15.186$\pm$0.012  &  14.434$\pm$0.014  &  14.192$\pm$0.019 &  2.$''$56  & 0.34 & late type & 5.1$^{+4.2}_{-1.5}$  \\ 
\hline
\hline 
\end{tabular}
\end{table*}

We obtained from {\itshape Swift/XRT} observations the most accurate source arcsecond-sized positional accuracy, which allowed us to perform a reliable search for counterparts from infrared to optical bands by using all of the available catalogues in the HEASARC database.

Four near-infrared (NIR) sources were detected by the UKIDSS Galactic Plane Survey (Lucas et al. 2008) inside the 99$\%$ confidence {\itshape XRT} error circle. Their main characteristics are listed in Table 3: the JHK magnitudes with lower limits derived according to Lawrence et al. (2007), offset from the {\itshape XRT} coordinates, Q value (see later in the section), inferred spectral type, and distance. Fig. 8 shows the {\itshape XRT} error circles  at 90$\%$ and 99$\%$ confidence levels, superimposed on the $J$ band UKIDSS image. We note that source n. 2 is very faint in all three NIR bands, while source n. 1 is very faint in the $J$ band and undetected in the $H$ and $K$ bands. As a consequence, we consider it very unlikely that both could be a reliable NIR counterpart of IGR J17419$-$2802. We are left with the two NIR sources, n. 3 and n. 4, which are bright and particularly close to the border of the 99$\%$ confidence error circle. In particular, we note that infrared source n. 4 (J174155.79-280157.9) is the brightest source and the only one also detected by \emph{Spitzer} in the mid-infrared (MIR) and by \emph{GAIA} Data Release 3 (DR3) in the optical band. Its MIR magnitude measurements are equal to $\sim$ 13.71 (3.6 $\mu$m) and 13.76 (4.5 $\mu$m), while no detection was obtained at wavelengths 5.8 $\mu$m and 8 $\mu$m. As for the optical band, J174155.79-280157.9 is listed in \emph{GAIA} DR3 (source ID 4060635740064278784) with magnitudes equal to  G=18.65, G$_{BP}$=20.28, and G$_{RP}$=17.48. 
We note that \emph{GAIA} mean G passband covers a wavelength range from the near ultraviolet (roughly 330 nm) to the NIR (roughly 1050 nm), while the other two passbands, G$_{BP}$ and G$_{RP}$, cover smaller wavelength ranges, from approximately 330 to 680 nm and 630 to 1050 nm, respectively. Two \emph{GAIA} distance estimates are available for J174155.79-280157.9 obtained with two different methods: geometric d$_{g}$=9.5$^{+2.2}_{-3.1}$ kpc (based only on the parallax) and photogeometric d$_{pg}$=5.1$^{+4.2}_{-1.5}$ kpc (which also uses the colour and the apparent magnitude of the star). We note that the nominal values of the distance from the two methods differ by a factor of $\sim$2, and accordingly the relative luminosity values would differ by a factor of $\sim$4. As for the choice of whether to use the geometric or photogeometric distance, we note that for stars with significantly large parallax uncertainties (which is the case for our specific source, i.e. 0.41$\pm$0.24 mas), photogeometric distances will generally be more precise and reliable than geometric ones (Bailer-Jones 2021). Therefore, according to the GAIA results, we adopted a distance of 5 kpc. 

If we use the reddening-free NIR diagnostic $Q$ of Negueruela \& Schurch (2007) for the sources from n. 2 to n. 4 in Table 3, then we find that none of them has a $Q$ value typical of early-type stars (i.e. $\leq$ 0). Conversely, they show $Q$ values typical of late-type K and M stars.

\begin{figure}
\begin{center}
\includegraphics[height=7.3 cm, angle=0]{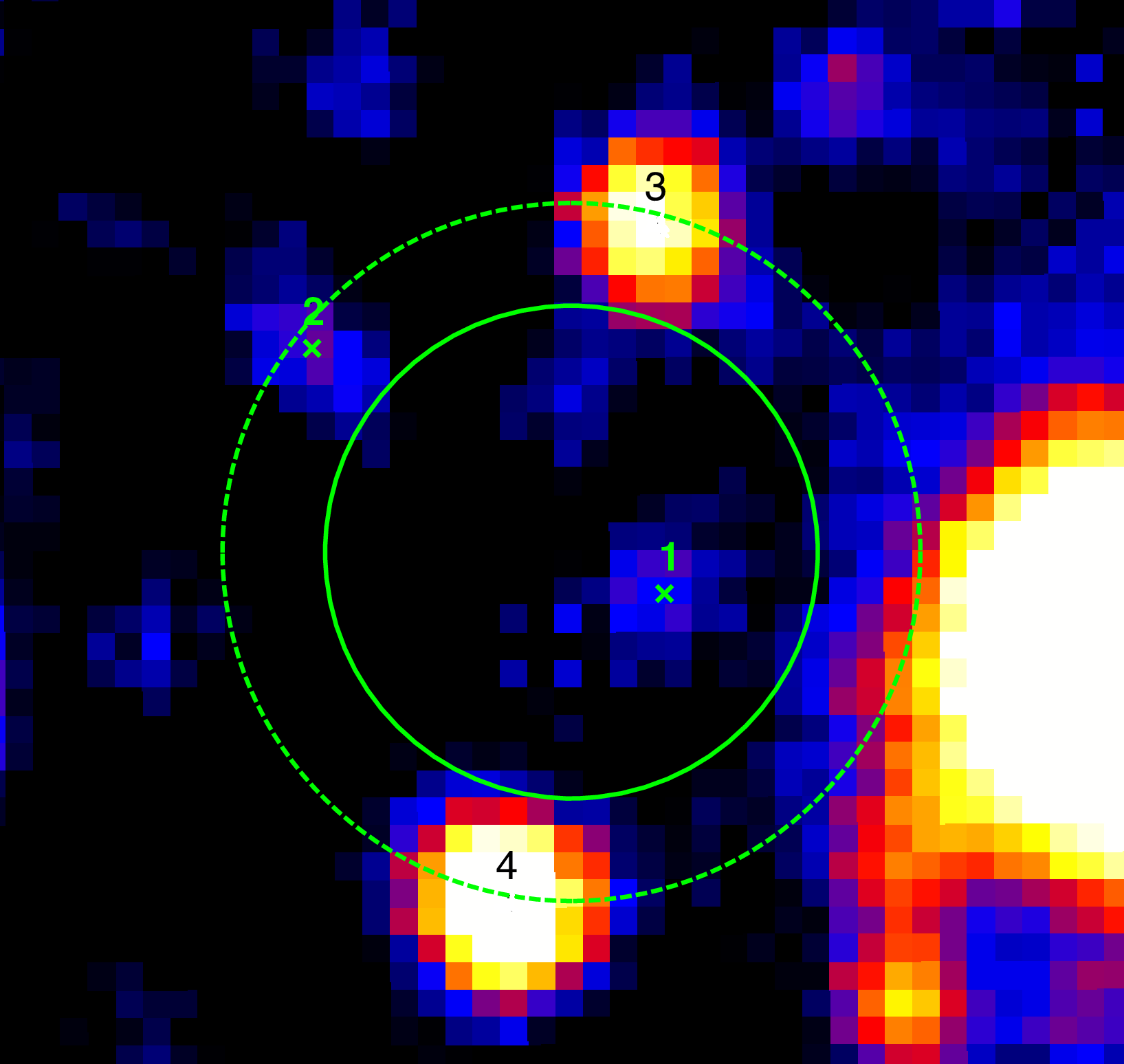}
\caption{UKIDSS image in the $J$ band (as downloaded from the UKIDSS archive, http://wsa.roe.ac.uk) with, superimposed, the {\itshape Swift/XRT} error circles at 90$\%$ confidence (solid green, radius of 1$''$.8) and 99$\%$ confidence (dashed green, radius of 2$''$.55).}
\end{center}
\end{figure}


\section{Summary and discussion}

We reported a broadband X-ray study of the hitherto poorly investigated unidentified X-ray transient \src. 
Three hard X-ray outbursts have been detected by \int\, with a typical 18--60 keV average (peak) X-ray flux of $\sim$1.0$\times$10$^{-10}$ erg cm$^{-2}$ s$^{-1}$ 
($\sim$3$\times$10$^{-10}$ erg cm$^{-2}$ s$^{-1}$). Their duration is constrained in the range 2-6 days, while their typical X-ray 
luminosity is of the order of $\sim$3$\times10^{35}$~erg~s$^{-1}$ (assuming source n. 4 in Table 3 as the best candidate counterpart). 
This indicates a peculiar short and faint X-ray transient nature for the source. The inferred duty cycle, as obtained from extensive \int\ observations above 18 keV, is $\sim$ 4$\%$.  The long-term light curve in the soft X-ray band (0.3--10 keV) clearly shows that the source spends most of the time undetected at very low X-ray fluxes (typical upper limit fluxes in the range 10$^{-12}$-10$^{-13}$ erg cm$^{-2}$ s$^{-1}$), the most stringent 3$\sigma$ flux (luminosity) upper limit being $<4.7\times10^{-14}$ erg cm$^{-2}$ s$^{-1}$ ($<1.4\times10^{32}$ erg s$^{-1}$). 
Conversely, the source has been detected only once, during \swift/XRT follow-up pointed observations that covered 
the decay phase of an X-ray outburst with unabsorbed flux (luminosity) of $\sim$6$\times10^{-11}$~erg~cm$^{-2}$~s$^{-1}$ ($\sim$1.8$\times$10$^{35}$ erg s$^{-1}$). The measured source dynamic range is $>$2,000 and $>$200 in the soft (0.3--10 keV) and hard (20--40 keV) X-ray band, respectively. We provide the most accurate X-ray position of the source to date by using \swift/XRT data. Two NIR counterparts are located inside the arcsecond-sized error circle; their photometric properties are compatible with a late-type spectral nature. Of the two, only the brightest one (n. 4 in Table 3) has also been detected by \emph{Spitzer} in the MIR and by \emph{GAIA} in the optical band, with an estimated distance of $\sim$ 5 kpc. We consider this infrared source as the best candidate counterpart of \src\ at lower energies, to date. 

All the collected multi-wavelength data can be used to investigate the possible nature of \src. 
In principle, the X-ray characteristics of the detected outbursts (e.g. duration of a very few days, high dynamic range, low duty cycle, and spectral shape) 
resemble those of supergiant fast X-ray transients (SFXTs; Sguera et al. 2005 and 2006; and Sidoli 2017 for a recent review), which are a subclass of high mass X-ray binaries (HMXBs). 
To date, about a dozen such objects have been reported in the literature.
Although classical SFXTs usually display above 20 keV hard X-ray outbursts that last much less than a day (i.e. a few hours), some of them are known to show longer hard X-ray activity, exceptionally lasting several days (e.g. IGR~J18483$-$0311, Sguera et al. 2015; IGR J17354$-$3255, Sguera et al. 2011, AX~J1949.8+2534, Sguera et al. 2017). These longer durations are comparable to those measured for the outbursts from \src. However, the SFXT interpretation for \src\ suffers a significant drawback when NIR data are taken into account. In fact, inside the source refined error circle we pinpointed two NIR counterparts whose magnitudes and colours are not compatible with being an early-type spectral star, which is strongly at odds with a SFXT nature. 
On the other hand, the two NIR counterparts, compatible with being late-type spectral stars, strongly suggest a low mass X--ray binary nature (LMXB). 
In this case, we note that the peculiar short duration of the reported three X-ray outbursts (i.e. a few days) is significantly shorter than the typical duration of 
classical transient LMXBs (of the order of weeks to months). In addition, their average X-ray luminosity ($\sim$3$\times10^{35}$~erg~s$^{-1}$) is one to three 
 orders of magnitude lower than typical luminous X-ray outbursts from classical LMXBs (up to $\sim$ 10$^{37-38}$~erg~s$^{-1}$).

 However, in this respect we note that in the last few decades, thanks to the improvement of the sensitivity of the last generation of X-ray detectors, 
 a significant growing number of LMXBs have been detected during faint outbursts reaching a much lower peak X-ray luminosity (2--10 keV) of $\sim$ 10$^{34-36}$~erg~s$^{-1}$. Because of this peculiar characteristic, they form a subclass of LMXBs that has been named very faint X-ray binaries (VFXBs; see Degenaar \& Wijnands 2009, Wijnands et al. 2015). 
VFXBs are the faintest known X-ray accretors (more than $\sim$40 identified systems are known to date), and they are a non-homogeneous class of sources. In fact, 
 some VFXBs are known to display a persistent or quasi-persistent X-ray luminosity of 
 $\sim$10$^{34}$~erg~s$^{-1}$. In particular, it has been predicted that some of them could correspond to the intermediate activity level of transitional millisecond pulsars (Heinke et al. 2015). Conversely, most of the VFXBs are transients and hence named very faint X-ray transients (VFXTs; Wijnands 2006). Their typical peak and quiescent X-ray luminosities are $\sim$10$^{34-36}$~erg~s$^{-1}$ and 10$^{30-33}$~erg~s$^{-1}$, respectively. The outburst duration of VFXTs is extremely variable (from several days to years, but typically of the order of weeks to months), and duty cycle values vary significantly (from 2\%-50\%). The X-ray characteristics of \src\ (late spectral type donor star, low luminosity and short duration X-ray outbursts, duty cycle, spectral shape, and very low X-ray luminosity in quiescence) strongly resemble those of VFXTs. In addition, most of them have been detected in the Galactic centre region, as is the case of \src. Hence we propose \src\ as a new member of the VFXT class. 
 Detailed spectral and timing investigations of their outbursts above 20 keV are particularly rare (e.g. Koch et al. 2014, Barlow et al. 2005). They are important since they allow us to probe the poorly known temporal and spectral characteristics of the hard accretion state in these systems. In this work, for the first time, the entire outburst activity of a VFXT above 20 keV was investigated in detail, from quiescence to outburst and then back to quiescence. In particular, our constrained outburst duration (typically a few days) is the shortest ever measured for an outburst above 20 keV (that we are aware of). In fact, previously reported constrained short-duration outbursts above 20 keV were of the order of over two weeks (e.g. Koch et al. 2014). Conversely, typical outburst durations of VFXTs in the soft band (0.3-10 keV) are longer, that is, of the order of weeks. We argue that the peculiar short duration of the reported outbursts detected by \inte\ could be due to their relatively hard X-ray faintness. In fact, only the brightest and shortest peak of the outbursts is detectable by \textrm{INTEGRAL}, while the longer and lower intensity X-ray outburst activity is just too faint to be detected at energies above 20 keV, being below the satellite hard X-ray sensitivity. 
 
VFXTs are an intriguing class of sources to study. Little is known about the mechanism behind their very low accretion rate, and they could represent a gateway to exploring accretion in a poorly studied mass-accretion regime. Furthermore, their very low time-averaged mass-accretion rates are problematic to explain in standard binary evolution models, as discussed in detail by King \& Wijnands (2006). We propose \src\ as a new member of this still small class of transients. NIR and optical spectroscopy follow-ups of the pinpointed putative late-type companion donor star are needed to fully establish its spectral characteristics and confirm our proposed VFXT nature. 


\begin{acknowledgements}
We thank the referee for the prompt and constructive comments. 
This work is based on observations performed with \xmm, \chandra\ , \swift\ and \int\ satellites. 
We made use of the High Energy Astrophysics Science Archive Research Center (HEASARC), a service of the Astrophysics Science Division at NASA/GSFC.
This work has made use of data from the European Space Agency (ESA) mission Gaia (https://www.cosmos.esa.int/gaia), processed by the Gaia Data Processing and Analysis Consortium (DPAC, https://www.cosmos.esa.int/web/gaia/dpac/consortium). 
We acknowledge funding from the grant entitled ``Bando Ricerca Fondamentale INAF 2023". 

\end{acknowledgements}



%

\clearpage

\begin{appendix} 

\section{Summary of the soft X-ray observations and upper limits}
We report in Table A.1 the summary of the soft X-ray observations and the relative upper limits. T$_{exp}$ is the nominal exposure time of the observation. The last column list the flux corrected for the absorption (0.3-10 keV). 
Upper limits (3$\sigma$) and detections were estimated assuming a power law model ($\Gamma$=2) and N$_{\rm H} = 1\times10^{22}$cm$^{-2}$.


\begin{center}
\begin{table*}
{\small 
\caption{Summary of the soft X-ray observations.}
\label{tab:log}
\hfill{}
\begin{tabular}{rllclcl}
\hline
Date       &  Date    & Instrument  & ObsID   & T$_{exp}$  & Off-axis angle &  Unabs.Flux  \\  
(yyyy-mm-dd)   &  (MJD)     &        &      &  (ks)   &   (arcmin)  &  (erg cm$^{-2}$ s$^{-1}$)  \\
\hline
2006-02-27    & 53793.873  & \xmm\ (Slew) &   $-$  &  $-$    &  $-$     &  $<7.2\times10^{-12}$  \\  
2006-03-08    & 53802.935  & \xmm\ (Slew) &   $-$  &  $-$    &  $-$     &  $<1.1\times10^{-11}$  \\  
2006-09-20    & 53998.806  & \xmm\ (Slew) &   $-$  &  $-$    &  $-$     &  $<3.3\times10^{-12}$ \\  
2012-03-01    & 55987.272  & \xmm\ (Slew) &   $-$  &  $-$    &  $-$     &  $<8.0\times10^{-12}$ \\ 
2018-03-03    & 58180.985  & \xmm\ (Slew) &   $-$  &  $-$    &  $-$     &  $<1.9\times10^{-11}$  \\ 
2017-09-05   & 58001.683  &  \xmm\    &  0801680101  & 28    &  10    & $<5.0\times10^{-14}$ \\ 
2005-10-03   & 53646.701  &  \sw/XRT   &  00030330001  & 1.8   &  0.7    & $(6.2\pm{0.2})\times10^{-11}$ \\ 
2005-10-11   & 53654.098  &  \sw/XRT   &  00030330003  & 0.7   &  1.5    &  $<1.0\times10^{-12}$ \\ 
2005-10-13   & 53656.344  &  \sw/XRT   &  00030330004  & 1.3   &  2.1    &  $<1.6\times10^{-12}$ \\ 
2018-02-15   & 58164.316  &  \sw/XRT   &  00093619001  & 0.1   &  10    &  $<7.9\times10^{-12}$ \\  
2018-03-01   & 58178.074  &  \sw/XRT   &  00093620002  & 0.1   &  12    &  $<8.1\times10^{-12}$ \\  
2018-03-29   & 58206.160  &  \sw/XRT   &  00093619004  & 0.1   &  11    &  $<8.6\times10^{-12}$ \\  
2018-03-29   & 58206.164  &  \sw/XRT   &  00093620004  & 0.1   &  13    &  $<1.7\times10^{-10}$ \\  
2019-04-04   & 58577.195  &  \sw/XRT   &  00095243001  & 0.1   &  9.6    &  $<7.9\times10^{-12}$ \\ 
2019-04-18   & 58591.086  &  \sw/XRT   &  00095243002  & 0.1   &  9.8    &  $<8.4\times10^{-12}$ \\  
2019-05-09   & 58612.391  &  \sw/XRT   &  00095243004  & 0.1   &  9.8    &  $<1.1\times10^{-11}$ \\  
2019-05-16   & 58619.617  &  \sw/XRT   &  00095243005  & 0.1   &  9.7    &  $<8.3\times10^{-12}$ \\  
2019-05-30   & 58633.109  &  \sw/XRT   &  00095243006  & 0.1   &  9.4    &  $<7.6\times10^{-12}$ \\   
2019-07-11   & 58675.273  &  \sw/XRT   &  00095243009  & 0.1   &  9.7    &  $<7.7\times10^{-12}$ \\  
2019-07-25   & 58689.355  &  \sw/XRT   &  00095243010  & 0.1   &  9.8    &  $<8.0\times10^{-12}$ \\  
2019-08-08   & 58703.293  &  \sw/XRT   &  00095243011  & 0.1   &  9.4    &  $<1.2\times10^{-11}$ \\  
2019-08-22   & 58717.180  &  \sw/XRT   &  00095243012  & 0.1   &  9.4    &  $<7.3\times10^{-12}$ \\  
2019-09-05   & 58731.309  &  \sw/XRT   &  00095243013  & 0.1   &  9.6    &  $<7.7\times10^{-12}$ \\  
2019-09-19   & 58745.328  &  \sw/XRT   &  00095243014  & 0.1   &  9.5    &  $<1.0\times10^{-11}$ \\  
2019-10-17   & 58773.113  &  \sw/XRT   &  00095243015  & 0.1   &  9.7    &  $<1.5\times10^{-11}$ \\ 
2020-02-13   & 58892.199  &  \sw/XRT   &  00095243016  & 0.1   &  9.8    &  $<8.0\times10^{-12}$ \\ 
2020-02-27   & 58906.355  &  \sw/XRT   &  00095243017  & 0.1   &  9.9    &  $<7.1\times10^{-12}$ \\  
2020-03-12   & 58920.691  &  \sw/XRT   &  00095243018  & 0.1   &  9.8    &  $<7.8\times10^{-12}$ \\  
2020-03-26   & 58934.348  &  \sw/XRT   &  00095243021  & 0.1   &  9.6    &  $<1.1\times10^{-11}$ \\  
2020-04-16   & 58955.074  &  \sw/XRT   &  00095243022  & 0.1   &  9.5    &  $<7.6\times10^{-12}$ \\     
2020-04-30   & 58969.160  &  \sw/XRT   &  00095243023  & 0.1   &  9.3    &  $<1.1\times10^{-11}$ \\   
2020-05-14   & 58983.168  &  \sw/XRT   &  00095243024  & 0.1   &  9.6    &  $<7.8\times10^{-12}$ \\
2021-04-08   & 59312.902  &  \sw/XRT   &  00096426001  & 0.1   &  9.6    &  $<8.3\times10^{-12}$ \\
2021-04-24   & 59328.973  &  \sw/XRT   &  00096426002  & 0.1   &  10     &  $<8.0\times10^{-12}$ \\ 
2021-05-21   & 59355.527  &  \sw/XRT   &  00096426003  & 0.1   &  10     &  $<7.8\times10^{-12}$ \\ 
2021-06-18   & 59383.879  &  \sw/XRT   &  00096426004  & 0.1   &  9.8    &  $<8.0\times10^{-12}$ \\  
2021-07-15   & 59410.820  &  \sw/XRT   &  00096426005  & 0.1   &  9.8    &  $<8.0\times10^{-12}$ \\  
2021-07-29   & 59424.898  &  \sw/XRT   &  00096426006  & 0.1   &  9.8    &  $<7.9\times10^{-12}$ \\  
2021-08-14   & 59440.621  &  \sw/XRT   &  00096426007  & 0.1   &  9.9    &  $<7.5\times10^{-12}$ \\  
2021-08-28   & 59454.828  &  \sw/XRT   &  00096426008  & 0.1   &  10     &  $<7.6\times10^{-12}$ \\     
2021-09-11   & 59468.695  &  \sw/XRT   & 	 00096426009  & 0.2   &  10     &  $<4.7\times10^{-12}$ \\   
2021-10-07   & 59494.652  &  \sw/XRT   &  00096426012  & 0.2   &  9.8    &  $<5.6\times10^{-12}$ \\  
2021-10-21   & 59508.719  &  \sw/XRT   &  00096426013  & 0.1   &  9.8    &  $<8.0\times10^{-12}$ \\ 
2022-04-14   & 59683.488  &  \sw/XRT   &  00096913001  & 0.1   &  7.3    &  $<8.8\times10^{-12}$ \\ 
2022-05-04   & 59703.840  &  \sw/XRT   &  00096913004  & 0.1   &  7.2    &  $<7.0\times10^{-12}$ \\    
2022-05-12   & 59711.348  &  \sw/XRT   &  00096913005  & 0.1   &  7.5    &  $<8.6\times10^{-12}$ \\    
2022-07-21   & 59781.375  &  \sw/XRT   &  00096913009  & 0.1   &  7.4    &  $<7.3\times10^{-12}$ \\ 
2022-08-04   & 59795.570  &  \sw/XRT   &  00096913010  & 0.1   &  7.5    &  $<7.4\times10^{-12}$ \\ 
2022-08-18   & 59809.223  &  \sw/XRT   &  00096913011  & 0.1   &  7.6    &  $<7.8\times10^{-12}$ \\  
2022-09-15   & 59837.461  &  \sw/XRT   &  00096913013  & 0.1   &  7.5    &  $<8.1\times10^{-12}$ \\  
2022-09-28   & 59850.262  &  \sw/XRT   &  00096913014  & 0.1   &  7.6    &  $<7.6\times10^{-12}$ \\  
2022-10-13   & 59865.438  &  \sw/XRT   &  00096913015  & 0.1   &  7.7    &  $<8.4\times10^{-12}$ \\ 
2022-10-26   & 59878.297  &  \sw/XRT   &  00096913016  & 0.1   &  7.5    &  $<7.6\times10^{-12}$ \\ 
2023-02-09   & 59984.223  &  \sw/XRT   &  00096913017  & 0.1   &  7.5    &  $<7.8\times10^{-12}$ \\ 
2023-02-23   & 59998.199  &  \sw/XRT   &  00096913018  & 0.1   &  7.6    &  $<7.7\times10^{-12}$ \\  
2007-07-30   & 54311.612  & \chandra/ACIS-I & 8202      &  14   &  9.8    &  $<4.7\times10^{-14}$ \\     
2008-05-16   & 54602.706  & \chandra/ACIS-I & 8643      &  2.1   &  5.9    &  $<1.1\times10^{-13}$ \\  
2008-05-16   & 54602.735  & \chandra/ACIS-I & 8654      &  2.1   &  8.6    &  $<1.1\times10^{-13}$ \\    
2009-08-03   & 55046.003  & \chandra/ACIS-I & 10689      &  0.9   &  0.1    &  $<2.2\times10^{-13}$ \\  
2016-05-11   & 57519.224  & \chandra/ACIS-I & 18312      &  1.9   &  9.1    &  $<2.2\times10^{-13}$ \\ 
2016-05-11   & 57519.255  & \chandra/ACIS-I & 18313      &  1.9   &  12     &  $<6.0\times10^{-13}$ \\   
2016-05-11   & 57519.318  & \chandra/ACIS-I & 18315      &  1.9   &  6.1  &  $<1.3\times10^{-13}$ \\   
\hline
\end{tabular}}
\hfill{}       
\end{table*}
\end{center}

\end{appendix}

\end{document}